\documentclass[%
 reprint,
superscriptaddress,
 amsmath,amssymb,
 aps,
]{revtex4-2}

\usepackage{graphicx}
\usepackage{dcolumn}
\usepackage{bm}
\usepackage{multirow}
\usepackage{color}

\begin{document}


\title{First results from the search for an excess of $\bar{\nu}_{e}$ events in JSNS$^2$}

\author{D. H. Lee}
\email{leedh@post.kek.jp}
\affiliation{High Energy Accelerator Research Organization (KEK), 1-1 Oho, Tsukuba, Ibaraki, 305-0801, Japan}
\author{S. Ajimura}
\affiliation{Research Center for Nuclear Physics, Osaka University, 10-1 Mihogaoka, Ibaraki, Osaka, 567-0047, Japan}
\author{A. Antonakis}
\affiliation{University of Michigan, 500 S. State Street, Ann Arbor, Michigan 48109, USA}
\author{M. Botran}
\affiliation{University of Michigan, 500 S. State Street, Ann Arbor, Michigan 48109, USA}
\author{M. K. Cheoun}
\affiliation{Department of Physics and OMEG Institute, Soongsil University, 369 Sangdo-ro, Dongjak-gu, Seoul, 06978, Korea}
\author{J. H. Choi}
\affiliation{Laboratory for High Energy Physics, Dongshin University, 67, Dongshindae-gil, Naju-si, Jeollanam-do, 58245, Korea}
\author{J. W. Choi}
\affiliation{Department of Physics, Chonnam National University, 77, Yongbong-ro, Buk-gu, Gwangju, 61186, Korea}
\author{J. Y. Choi}
\affiliation{Department of Fire Safety, Seoyeong University, 1 Seogang-ro, Buk-gu, Gwangju, 61268, Korea} 
\author{\\ T. Dodo}
\affiliation{Research Center for Neutrino Science, Tohoku University, 6-3 Azaaoba, Aramaki, Aoba-ku, Sendai, Miyagi 980-8578, Japan}
\affiliation{Advanced Science Research Center, JAEA, 2-4 Shirakata, Tokai-mura, Naka-gun, Ibaraki 319-1195, Japan}
\author{H. Furuta}
\affiliation{Research Center for Neutrino Science, Tohoku University, 6-3 Azaaoba, Aramaki, Aoba-ku, Sendai, Miyagi 980-8578, Japan}
\author{J. H. Goh}
\affiliation{Department of Physics, Kyung Hee University, 26, Kyungheedae-ro, Dongdaemun-gu, Seoul 02447, Korea}
\author{M. Harada}
\affiliation{J-PARC Center, JAEA, 2-4 Shirakata, Tokai-mura, Naka-gun, Ibaraki 319-1195, Japan}
\author{S. Hasegawa}
\affiliation{Advanced Science Research Center, JAEA, 2-4 Shirakata, Tokai-mura, Naka-gun, Ibaraki 319-1195, Japan}
\affiliation{J-PARC Center, JAEA, 2-4 Shirakata, Tokai-mura, Naka-gun, Ibaraki 319-1195, Japan}
\author{Y. Hino} \thanks{Now at KEK}
\affiliation{Research Center for Neutrino Science, Tohoku University, 6-3 Azaaoba, Aramaki, Aoba-ku, Sendai, Miyagi 980-8578, Japan}
\author{T. Hiraiwa} \thanks{Now at RIKEN}
\affiliation{Research Center for Nuclear Physics, Osaka University, 10-1 Mihogaoka, Ibaraki, Osaka, 567-0047, Japan}
\author{W. S. Hwang}
\affiliation{Department of Physics, Kyung Hee University, 26, Kyungheedae-ro, Dongdaemun-gu, Seoul 02447, Korea}
\author{\\ T. Iida}
\affiliation{Faculty of Pure and Applied Sciences, University of Tsukuba,\\ Tennodai 1-1-1, Tsukuba, Ibaraki, 305-8571, Japan}
\author{E. Iwai} \thanks{Now at RIKEN}
\affiliation{University of Michigan, 500 S. State Street, Ann Arbor, Michigan 48109, USA}
\author{S. Iwata} \thanks{Now at Tokyo Metropolitan College of Industrial Technology (Tokyo Metro. Col. of Indus. Tech.)}
\affiliation{Department of Physics, Kitasato University, 1 Chome-15-1 Kitazato, Minami Ward, Sagamihara, Kanagawa, 252-0329, Japan}
\author{H. I. Jang}
\affiliation{Department of Fire Safety, Seoyeong University, 1 Seogang-ro, Buk-gu, Gwangju, 61268, Korea}
\author{J. S. Jang}
\affiliation{GIST College, Gwangju Institute of Science and Technology, 123 Cheomdangwagi-ro, Buk-gu, Gwangju, 61005, Korea}
\author{M. C. Jang}
\affiliation{Department of Physics, Chonnam National University, 77, Yongbong-ro, Buk-gu, Gwangju, 61186, Korea}
\author{H. K. Jeon}
\affiliation{Department of Physics, Sungkyunkwan University, 2066, Seobu-ro, Jangan-gu, Suwon-si, Gyeonggi-do, 16419, Korea}
\author{S. H. Jeon}
\affiliation{Laboratory for High Energy Physics, Dongshin University, 67, Dongshindae-gil, Naju-si, Jeollanam-do, 58245, Korea}
\author{K. K. Joo}
\affiliation{Department of Physics, Chonnam National University, 77, Yongbong-ro, Buk-gu, Gwangju, 61186, Korea}
\author{\\ D. E. Jung}
\affiliation{Department of Physics, Chonnam National University, 77, Yongbong-ro, Buk-gu, Gwangju, 61186, Korea}
\author{S. K. Kang}
\affiliation{School of Liberal Arts, Seoul National University of Science and Technology, 232 Gongneung-ro, Nowon-gu, Seoul, 139-743, Korea}
\author{Y. Kasugai}
\affiliation{J-PARC Center, JAEA, 2-4 Shirakata, Tokai-mura, Naka-gun, Ibaraki 319-1195, Japan}
\author{T. Kawasaki}%
\affiliation{Department of Physics, Kitasato University, 1 Chome-15-1 Kitazato, Minami Ward, Sagamihara, Kanagawa, 252-0329, Japan}
\author{E. M. Kim}
\affiliation{Department of Physics, Chonnam National University, 77, Yongbong-ro, Buk-gu, Gwangju, 61186, Korea}
\author{E. J. Kim}
\affiliation{Division of Science Education, Jeonbuk National University, 567 Baekje-daero, Deokjin-gu, Jeonju-si, Jeollabuk-do, 54896, Korea}
\author{J. Y. Kim}
\affiliation{Department of Physics, Chonnam National University, 77, Yongbong-ro, Buk-gu, Gwangju, 61186, Korea}
\author{S. Y. Kim}
\affiliation{Department of Physics, Chonnam National University, 77, Yongbong-ro, Buk-gu, Gwangju, 61186, Korea}
\author{\\ S. B. Kim}
\affiliation{School of Physics, Sun Yat-sen (Zhongshan) University, Haizhu District, Guangzhou, 510275, China}
\author{W. Kim}
\affiliation{Department of Physics, Kyungpook National University, 80 Daehak-ro, Buk-gu, Daegu, 41566, Korea}
\author{H. Kinoshita}
\affiliation{J-PARC Center, JAEA, 2-4 Shirakata, Tokai-mura, Naka-gun, Ibaraki 319-1195, Japan}
\author{T. Konno}
\affiliation{Department of Physics, Kitasato University, 1 Chome-15-1 Kitazato, Minami Ward, Sagamihara, Kanagawa, 252-0329, Japan}
\author{K. Kuwata}
\affiliation{Research Center for Neutrino Science, Tohoku University, 6-3 Azaaoba, Aramaki, Aoba-ku, Sendai, Miyagi 980-8578, Japan}
\author{S. Lee}
\affiliation{Department of Physics, Kyung Hee University, 26, Kyungheedae-ro, Dongdaemun-gu, Seoul 02447, Korea}
\author{I. T. Lim}
\affiliation{Department of Physics, Chonnam National University, 77, Yongbong-ro, Buk-gu, Gwangju, 61186, Korea}
\author{C. Little}
\affiliation{University of Michigan, 500 S. State Street, Ann Arbor, Michigan 48109, USA}
\author{T. Maruyama}
\affiliation{High Energy Accelerator Research Organization (KEK), 1-1 Oho, Tsukuba, Ibaraki, 305-0801, Japan}
\author{E. Marzec}
\affiliation{University of Michigan, 500 S. State Street, Ann Arbor, Michigan 48109, USA}
\author{S. Masuda}
\affiliation{J-PARC Center, JAEA, 2-4 Shirakata, Tokai-mura, Naka-gun, Ibaraki 319-1195, Japan}
\author{S. Meigo}
\affiliation{J-PARC Center, JAEA, 2-4 Shirakata, Tokai-mura, Naka-gun, Ibaraki 319-1195, Japan}
\author{S. Monjushiro}
\affiliation{High Energy Accelerator Research Organization (KEK), 1-1 Oho, Tsukuba, Ibaraki, 305-0801, Japan}
\author{D. H. Moon}
\affiliation{Department of Physics, Chonnam National University, 77, Yongbong-ro, Buk-gu, Gwangju, 61186, Korea}
\author{T. Nakano}
\affiliation{Research Center for Nuclear Physics, Osaka University, 10-1 Mihogaoka, Ibaraki, Osaka, 567-0047, Japan}
\author{M. Niiyama}
\affiliation{Department of Physics, Kyoto Sangyo University, Motoyama, Kamigamo, Kita-Ku, Kyoto-City, 603-8555, Japan}
\author{K. Nishikawa} \thanks{Deceased}
\affiliation{High Energy Accelerator Research Organization (KEK), 1-1 Oho, Tsukuba, Ibaraki, 305-0801, Japan}
\author{M. Noumachi}
\affiliation{Research Center for Nuclear Physics, Osaka University, 10-1 Mihogaoka, Ibaraki, Osaka, 567-0047, Japan}
\author{M. Y. Pac}
\affiliation{Laboratory for High Energy Physics, Dongshin University, 67, Dongshindae-gil, Naju-si, Jeollanam-do, 58245, Korea}
\author{B. J. Park}
\affiliation{Department of Physics, Kyungpook National University, 80 Daehak-ro, Buk-gu, Daegu, 41566, Korea}
\author{H. W. Park}
\affiliation{Department of Physics, Chonnam National University, 77, Yongbong-ro, Buk-gu, Gwangju, 61186, Korea}
\author{J. B. Park}
\affiliation{Department of Physics and OMEG Institute, Soongsil University, 369 Sangdo-ro, Dongjak-gu, Seoul, 06978, Korea}
\author{Jisu Park}
\affiliation{Department of Physics, Chonnam National University, 77, Yongbong-ro, Buk-gu, Gwangju, 61186, Korea}
\author{J. S. Park}
\affiliation{Department of Physics, Kyungpook National University, 80 Daehak-ro, Buk-gu, Daegu, 41566, Korea}
\author{R. G. Park}
\affiliation{Department of Physics, Chonnam National University, 77, Yongbong-ro, Buk-gu, Gwangju, 61186, Korea}
\author{\\ S. J. M. Peeters}
\affiliation{Department of Physics and Astronomy, University of Sussex, Falmer, Brighton, BN1 9RH, United Kingdom}
\author{G. Roellinghoff}
\affiliation{Department of Physics, Sungkyunkwan University, 2066, Seobu-ro, Jangan-gu, Suwon-si, Gyeonggi-do, 16419, Korea}
\author{C. Rott}
\affiliation{Department of Physics and Astronomy, University of Utah, 201 Presidents' Cir, Salt Lake City, UT 84112, U.S.A}
\author{J. W. Ryu}
\affiliation{Department of Physics, Kyungpook National University, 80 Daehak-ro, Buk-gu, Daegu, 41566, Korea}
\author{K. Sakai}
\affiliation{J-PARC Center, JAEA, 2-4 Shirakata, Tokai-mura, Naka-gun, Ibaraki 319-1195, Japan}
\author{S. Sakamoto}
\affiliation{J-PARC Center, JAEA, 2-4 Shirakata, Tokai-mura, Naka-gun, Ibaraki 319-1195, Japan}
\author{T. Shima}
\affiliation{Research Center for Nuclear Physics, Osaka University, 10-1 Mihogaoka, Ibaraki, Osaka, 567-0047, Japan}
\author{\\ C. D. Shin}
\affiliation{Laboratory for High Energy Physics, Dongshin University, 67, Dongshindae-gil, Naju-si, Jeollanam-do, 58245, Korea}
\author{J. Spitz}
\affiliation{University of Michigan, 500 S. State Street, Ann Arbor, Michigan 48109, USA}
\author{I. Stancu}
\affiliation{University of Alabama, Tuscaloosa, AL, 35487, USA}
\author{F. Suekane}
\affiliation{Research Center for Neutrino Science, Tohoku University, 6-3 Azaaoba, Aramaki, Aoba-ku, Sendai, Miyagi 980-8578, Japan}
\author{Y. Sugaya}
\affiliation{Research Center for Nuclear Physics, Osaka University, 10-1 Mihogaoka, Ibaraki, Osaka, 567-0047, Japan}
\author{K. Suzuya}
\affiliation{J-PARC Center, JAEA, 2-4 Shirakata, Tokai-mura, Naka-gun, Ibaraki 319-1195, Japan}
\author{M. Taira}
\affiliation{High Energy Accelerator Research Organization (KEK), 1-1 Oho, Tsukuba, Ibaraki, 305-0801, Japan}
\author{Y.Takeuchi}
\affiliation{Faculty of Pure and Applied Sciences, University of Tsukuba,\\ Tennodai 1-1-1, Tsukuba, Ibaraki, 305-8571, Japan}
\author{W. Wang}
\affiliation{School of Physics, Sun Yat-sen (Zhongshan) University, Haizhu District, Guangzhou, 510275, China}
\author{\\ J. Waterfield}
\affiliation{Department of Physics and Astronomy, University of Sussex, Falmer, Brighton, BN1 9RH, United Kingdom}
\author{W. Wei}
\affiliation{School of Physics, Sun Yat-sen (Zhongshan) University, Haizhu District, Guangzhou, 510275, China}
\author{R. White}
\affiliation{Department of Physics and Astronomy, University of Sussex, Falmer, Brighton, BN1 9RH, United Kingdom}
\author{Y. Yamaguchi}
\affiliation{J-PARC Center, JAEA, 2-4 Shirakata, Tokai-mura, Naka-gun, Ibaraki 319-1195, Japan}
\author{M. Yeh}
\affiliation{Brookhaven National Laboratory, Upton, NY 11973-5000, U.S.A.}
\author{I. S. Yeo}
\affiliation{Laboratory for High Energy Physics, Dongshin University, 67, Dongshindae-gil, Naju-si, Jeollanam-do, 58245, Korea}
\author{C. Yoo}
\affiliation{Department of Physics, Kyung Hee University, 26, Kyungheedae-ro, Dongdaemun-gu, Seoul 02447, Korea}
\author{I. Yu}
\affiliation{Department of Physics, Sungkyunkwan University, 2066, Seobu-ro, Jangan-gu, Suwon-si, Gyeonggi-do, 16419, Korea}
\author{A. Zohaib}
\affiliation{Department of Physics, Chonnam National University, 77, Yongbong-ro, Buk-gu, Gwangju, 61186, Korea}

\collaboration{JSNS$^2$ Collaboration}



\date{\today}

\begin{abstract}
The JSNS$^2$ (J-PARC Sterile Neutrino Search at the J-PARC Spallation Neutron Source) 
experiment at the Material and Life Science Facility (MLF) of J-PARC is designed to 
directly test an excess on $\bar{\nu}_{e}$ 
events which was indicated by LSND (Liquid Scintillator Neutrino Detector). 
The combination of a short-pulsed proton beam 
and a gadolinium-loaded liquid scintillator provides an excellent signal-to-noise 
ratio. In this article, we report the first results of a direct test based on data 
collected in 2022. After applying all event selection criteria, two events are observed, 
consistent with the expected background of 2.3$\pm$0.4 events. No excess of $\bar{\nu}_e$
events are seen in this report, however the expected number of events due to LSND anomaly is 
1.1$\pm$0.5, thus this result is not yet conclusive. 
Data taking has been ongoing since 2021 and will continue in future runs. 
In addition, a new far detector has recently been constructed for the second phase
experiment, JSNS$^2$-II, marking an important milestone toward forthcoming measurements.

\end{abstract}

\maketitle


\section{\label{sec:intro} Introduction }
\indent

JSNS$^2$~\cite{CITE:JSNS2proposal,CITE:JSNS2TDR} aims to directly test
an excess of $\bar{\nu}_{e}$ events which was indicated by LSND~\cite{CITE:LSND}, 
the so called LSND anomaly. 
LSND reported a significant excess of observed $\bar{\nu}_e$ events above background with the 
neutrinos produced by muon decay-at-rest and the liquid scintillator detector,
thus JSNS$^2$ employs a same neutrino source, target, and 
detection principle as LSND, while incorporating modern technologies such as 
Gd-loaded liquid scintillator (Gd-LS), 500 MHz flash ADCs (FADCs), and a 
short pulsed proton beam to perform a stringent test. 
This excess may be related to observations reported by other experiments~\cite{CITE:BEST, CITE:MiniBooNE2018, CITE:REACTOR}, such as the sterile neutrino hypothesis that induces 
short-baseline neutrino oscillations; 
the results from JSNS$^2$ will also provide further insight into these interpretations.

The JSNS$^2$ experiment employs a 50-ton liquid scintillator detector on the third 
floor of the MLF, at a baseline of 24 m from the neutrino source~\cite{CITE:JSNS2NIM}.
The neutrino source is a liquid-mercury target designed to withstand a 1 MW proton beam. Although originally constructed as a spallation neutron source for material and life science studies, the mercury target simultaneously produces (anti)neutrinos.
JSNS$^2$ utilizes (anti)neutrinos from muon decay at rest 
($\mu^{+} \to e^{+} + \nu_{e} + \bar{\nu}_{\mu}$) to test the anomaly.
Protons are accelerated to 3 GeV by the Rapid Cycling Synchrotron at J-PARC and 
impinge on the mercury target. The beam operates at 25 Hz with two 100 ns wide bunches 
separated by 600 ns, providing excellent temporal separation between the signal and backgrounds.
The designed 1 MW beam power was achieved in 2024. Using the 
$^{12}$C$ (\nu_{e},e^{-})^{12}$N$_{g.s.}$ reaction (CNgs), 
the number of positive muons produced per 
proton at the target is measured to be 0.48$\pm$0.17~\cite{CITE:JSNS2CNgs}, 
which is about an order of magnitude larger
than that of the KARMEN experiment~\cite{CITE:KARMEN2} owing to the MLF's higher proton 
beam energy.
The $\bar{\nu}_e$
signal is detected through the inverse beta decay (IBD) reaction, $\bar{\nu}_e + p \to e^{+} + n$, in the liquid scintillator.
The neutrino target scintillator contains 0.1\% Gd by mass, enabling enhanced neutron 
capture after thermalization, which results in a $\sim$8 MeV $\gamma$-rays cascade 
after a mean capture time of $\sim$30 $\mu$s.
The prompt signal from the positron and the delayed signal from Gd capture form a coincidence 
signature. The most severe background arises from cosmogenic neutrons, which can mimic this 
coincidence by their recoil protons and captured neutrons~\cite{CITE:EPJC}.
To suppress such backgrounds, 10\% di-isopropylnaphthalene (DIN, C$_{16}$H$_{20}$) was 
dissolved into the 
Gd-LS from 2021 to 2022. DIN enhances pulse-shape discrimination (PSD) performance 
for prompt candidates.
A dedicated PSD algorithm utilizing full waveform information and a log-likelihood ratio method was 
developed to further reduce neutron-induced backgrounds~\cite{CITE:JSNS2PSD}.

Commissioning data were collected in 2020~\cite{CITE:EPJC}, and physics data have been accumulated 
since 2021. Despite annual installation and dismantling of the detector and scintillator at MLF, 
no significant degradation of the scintillator performance has been observed. In total, JSNS$^2$ 
accumulated 5.14$\times10^{22}$ protons on target (POT) from 2021 to 2025.
This article reports results from the 2022 dataset (0.82$\times$10$^{22}$ POT), corresponding 
to the first period with 10\% DIN concentration in the Gd-LS.

\section{\label{sec:detector} JSNS$^2$ detector}
\indent
Reference~\cite{CITE:JSNS2NIM} provides a detailed description of the JSNS$^2$ detector; 
only the components relevant to this report are summarized here.

The cylindrical detector consists of three concentric layers of liquid scintillator. From the 
innermost to the outermost regions the three concentric layers are the inner target, then the surrounding gamma-catcher and the veto layer. The inner target is filled with 
Gd-loaded liquid scintillator (Gd-LS) with 10\% DIN, and  
both gamma-catcher and veto layers contain pure liquid scintillator (pure LS) which 
have neither Gd nor DIN. 
The target region
is contained in an acrylic vessel with a diameter of 3.2 m and a height of 2.5 m.
The gamma-catcher and veto layers each have a typical thickness of 25 cm and 
are optically separated by black acrylic plates.
All layers are housed within a stainless-steel tank with a diameter of 4.6 m and a height of 3.5 m~\cite{CITE:HINOss}. 
The target contains approximately 17 tons of Gd-LS, while the outer regions together 
hold about 33 tons of pure LS.
The base solvent of both Gd-LS and pure LS is linear alkylbenzene (LAB),
containing 3 g/L of 2,5-diphenyloxazole (PPO) as the primary fluor, and 15 mg/L (Gd-LS) and
30 mg/L (pureLS) of 1,4-bis(2-methylstyryl)benzene (bis-MSB) as a secondary wavelength shifter.
DIN (produced by Eljen company: EJ-309~\cite{CITE:EJ}) exhibits long-term stability suitable 
for extended physics runs owing to its high chemical purity.  
Particles with large energy loss per unit length (dE/dx), in particular in DIN-containing 
scintillator, 
produce scintillation light with broader timing profiles and longer decay tails than those with 
small dE/dx particles.
The typical PSD performance of the JSNS$^2$ detector is described in Ref.~\cite{CITE:JSNS2PSD}.
Using DIN and a dedicated likelihood-based PSD algorithm, the experiment 
achieved a neutron rejection efficiency of 95.0$\pm$0.2\%
while maintaining a Michel-electron identification efficiency of 
92.8$\pm$1.8\%.
To reconstruct events in the target and gamma-catcher regions from scintillation light, ninety-six 
10-inch Hamamatsu R7081 photomultiplier tubes (PMTs) are installed around the boundary 
between the gamma-catcher and veto layers.
An additional twenty-four PMTs are mounted in the veto region to detect incoming particles such 
as cosmic-ray muons entering the detector.

Signals from all 120 PMTs are digitized using 500 MHz, 8-bit FADCs~\cite{CITE:JSNS2DAQ}.
A coincidence trigger has been implemented to efficiently acquire both prompt 
and delayed signals from IBD events.
A prompt trigger is issued when the analog sum of the 96 PMT signals from the target and
gamma-catcher regions exceeds a threshold of approximately 200 mV ($\sim$5 MeV) 
within a 1.7–10 $\mu$s window after the beam start timing.
If a prompt trigger exists, a delayed trigger is generated when the analog sum exceeds 
about 70 mV ($\sim$2 MeV) within a time window of 25 ms after the prompt trigger.
The beam timing in our trigger system is defined by a signal sent directly 
from the radio-frequency module of J-PARC accelerator.

\section{Calibrations}
\indent
In selecting IBD candidates, the event energy and vertex are reconstructed 
using the JADE (JSNS$^2$ Analysis Development Environment) software
framework~\cite{CITE:Johnathon}, which employs a likelihood maximization algorithm.
Detector calibrations to validate JADE were carried out 
with one- and three-dimensional calibration systems using $^{252}$Cf sources 
providing a cascade of $\gamma$-rays totaling $\sim$8 MeV 
from neutron captures on Gd (n-Gd) 
and with cosmogenic Michel-electrons, whose endpoint energy is $\sim$
53 MeV~\cite{CITE:Cf}. JSNS$^2$ adopts a simple charge-prediction model for 
the PMTs, parameterized by the distance and zenith angle between the 
scintillation light source and the center of each PMT sphere, which provides reliable event 
reconstruction performance.

After calibration, the reconstructed energy scale varies by a few percent over time 
and across vertex positions. These variations are corrected using cosmogenic Michel-electrons 
and n-Gd capture events observed in physics data made by cosmogenic neutrons, 
and are properly accounted for this analysis.
The typical energy calibration using Michel-electrons is also shown in Ref.~\cite{CITE:JSNS2KDAR}. 
The systematic uncertainty on the energy scale after calibration is estimated to be 0.8\%.
The uncertainty on the fiducial volume, determined from the three-dimensional $^{252}$Cf 
calibration, is currently $\sim$20\%~\cite{CITE:Cf}.

\section{Event Selection}
\indent
The event selection is optimized to efficiently search for an excess of $\bar{\nu}_{e}$ IBD 
events. 
Table~\ref{tab:EventSelection} summarizes IBD event selection criteria, and their efficiencies 
and uncertainties, evaluated to detect $\bar{\nu}_e$ IBD events.
The fiducial volume for this analysis is specified as $r<$140 cm and $|z| < $100 cm, where the origin is set at the center of the target ($r = \sqrt{x^2 + y^2}$).
Further details are provided in the following.

\begin{table}[htb]
    \centering
    \caption{\label{tab:EventSelection}
IBD selection criteria, and their efficiencies and uncertainties evaluated for the LSND anomaly  are summarized. The efficiencies are determined using a combination of data and Monte Carlo simulation. Each efficiency is quote relative to the previous efficiencies, and the cumulative total is given at the end.}
    \vspace{3pt}
    \begin{tabular}{cc}\hline
        Requirement & Relative Efficiency (\%) \\\hline
        --Prompt Candidate--\\

        $20\le E_{\mathrm{p}} \le 60$~MeV     & 100.0 \\
        $2\le \Delta T_{beam-p}\le 10$~$\mu$s & 46.5$\pm$0.5  \\
        PSD & 87.2$\pm$9.1\\
                                                            &   \\

        --Delayed Candidate-- \\
        $7\le E_{\mathrm{d}}\le 12$~MeV      & 75.3$\pm$0.9 \\
        Beam neutron rejection & 94.1$\pm$0.1 \\ 
                                                            &   \\

        --IBD paired Candidate-- \\
        $\Delta T_{p-d} \le 100 ~\mu$s & 96.6$^{+3.4}_{-5.6}$\\
        $\Delta$VTX$_{p-d} \le 60$~cm & 83.4$\pm$10.0 \\ 
        Background rejection likelihood & 70.5$\pm$1.5 \\ 
                                                            &     \\
                                                            
         --Muon and Michel electron rejections-- \\
        Muon rejection                       & 92.8$\pm$0.5 \\
        Michel electron rejections (veto layer)    & 97.0$\pm$0.03\\ 
        Michel electron rejections (baseline) & 86.1$\pm$1.4 \\
        Chimney passing muon rejections & 98.7$\pm$0.1 \\\hline 
        
        Cumulative total efficiency   & 12.4$^{+2.1}_{-2.2}$ \\\hline
    \end{tabular}
\end{table}

\subsection{IBD Prompt Candidates}
\indent
The IBD prompt-energy selection is identical to that of LSND 
to have a model-independent test. The selection efficiency, 
$20\le E_{\mathrm{p}} \le 60$~MeV, including the detector's trigger efficiency, is 100\%.
However, under a sterile neutrino hypothesis, the energy selection’s efficiency is 
sensitive to oscillation parameters, thus this dependency is described and consider 
further later.

\begin{figure}[htbp]
	\centering
	\begin{minipage}[b]{0.46\textwidth}
       \includegraphics[width=0.924\textwidth]{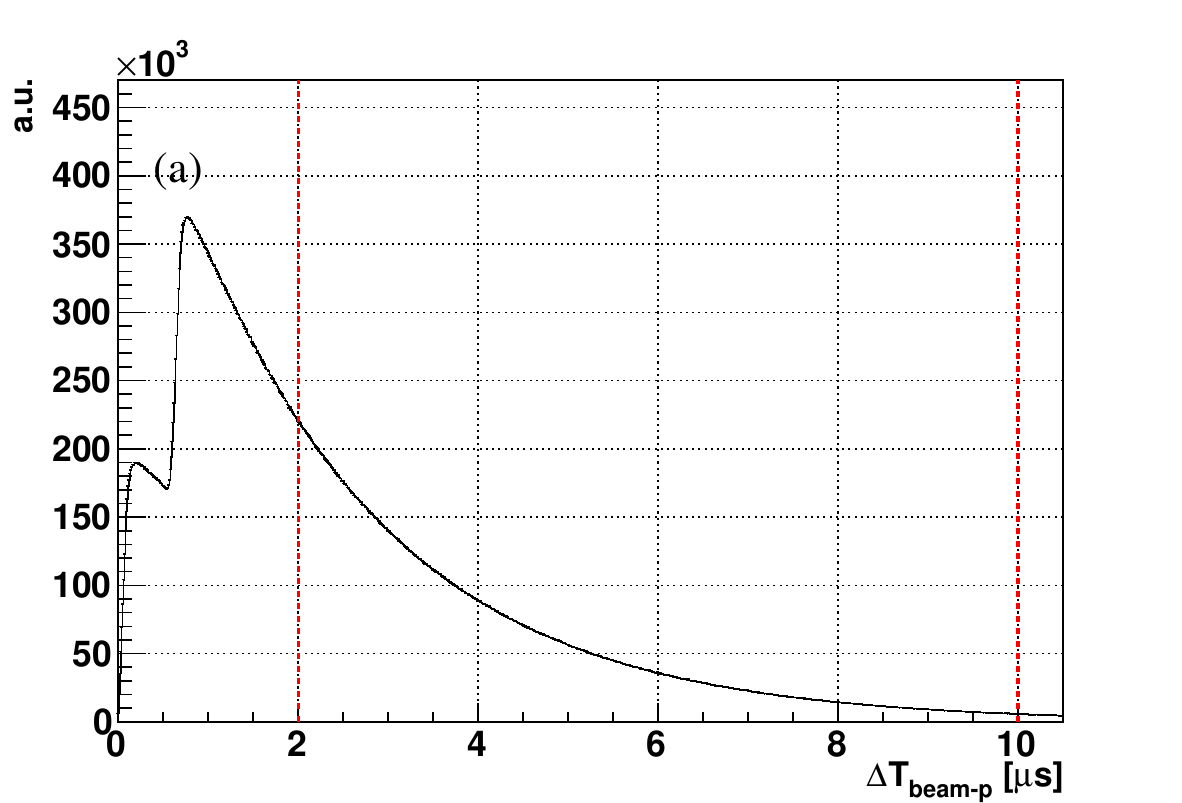}
			\qquad
    \end{minipage}       
 	\begin{minipage}[b]{0.46\textwidth}
       \includegraphics[width=0.924\textwidth]{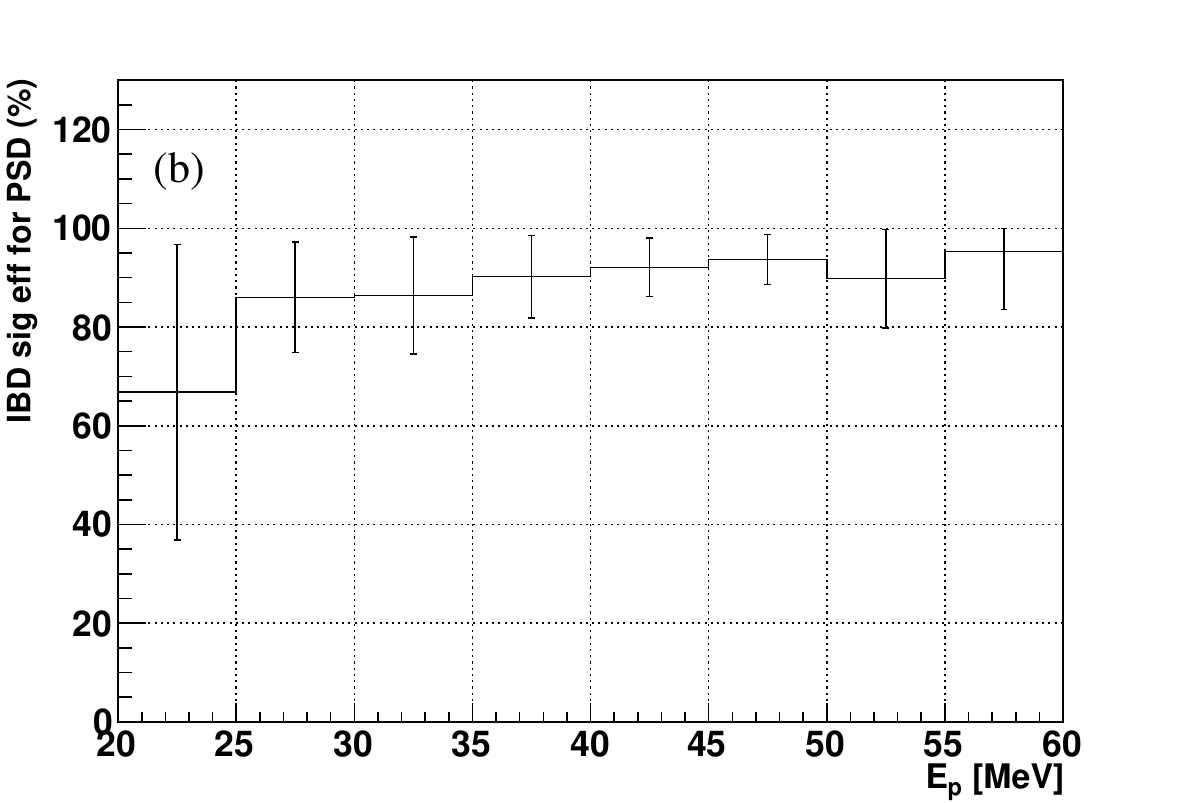}
			\qquad
    \end{minipage}     
	\caption{
    (a) Expected timing distribution of IBD prompt candidates relative to the beam start, obtained from MC simulation.
    (b) PSD selection efficiency for IBD events as a function of energy.
    The vertical axes of (a) is in arbitrary units.}    
\label{Fig:Prompt}
\end{figure}

The timing of the IBD prompt candidates relative to the beam pulse, $\Delta T_{beam-p}$, is required to 
satisfy $2\le \Delta T_{beam-p}\le 10$~$\mu$s. When the proton beam with two bunches impinges on the mercury 
target, muons are produced with a lifetime of 2.2 $\mu$s. Consequently, the expected arrival time 
of the $\bar{\nu}_{e}$ events in the JSNS$^2$ detector with respect to the beam start 
is shown in Fig.~\ref{Fig:Prompt}(a). 
This 8 $\mu$s time window strongly suppresses the cosmogenic backgrounds. In addition, 
events with $\Delta T_{beam-p}\le 2$~$\mu$s are rejected to suppress backgrounds from beam 
neutrons and neutrinos produced by $\pi$ or K decays. 
The clocks of FADCs are synchronized with the 40 ms 
accelerator repetition cycle, and the uncertainty associated with the FADC timing is negligible. 
The dominant systematic uncertainty for this selection arises from the trigger dead-time stability, estimated to be approximately 1\%.

The PSD capability
 was originally evaluated using cosmogenic Michel-electrons as a control sample~\cite{CITE:JSNS2PSD}. 
 However, subsequent studies revealed baseline shifts in the FADC readout following muon events
 due to electronics effects~\cite{CITE:baseline}, rendering Michel-electrons unsuitable for PSD 
 evaluation in IBD analyses. Instead, a $\gamma$-ray control sample is used to determine 
 the PSD performance. Figure~\ref{Fig:Prompt}(b) shows the updated PSD efficiency as a function 
 of energy while requiring approximately 99.7\% rejection of cosmogenic neutrons used in this manuscript. Uncertainties are derived from differences between the methods used to evaluate this efficiency. 

\subsection{IBD Delayed Candidates}
\begin{figure}[htbp]
	\centering
	\begin{minipage}[b]{0.49\textwidth}
      \includegraphics[width=0.98\textwidth]{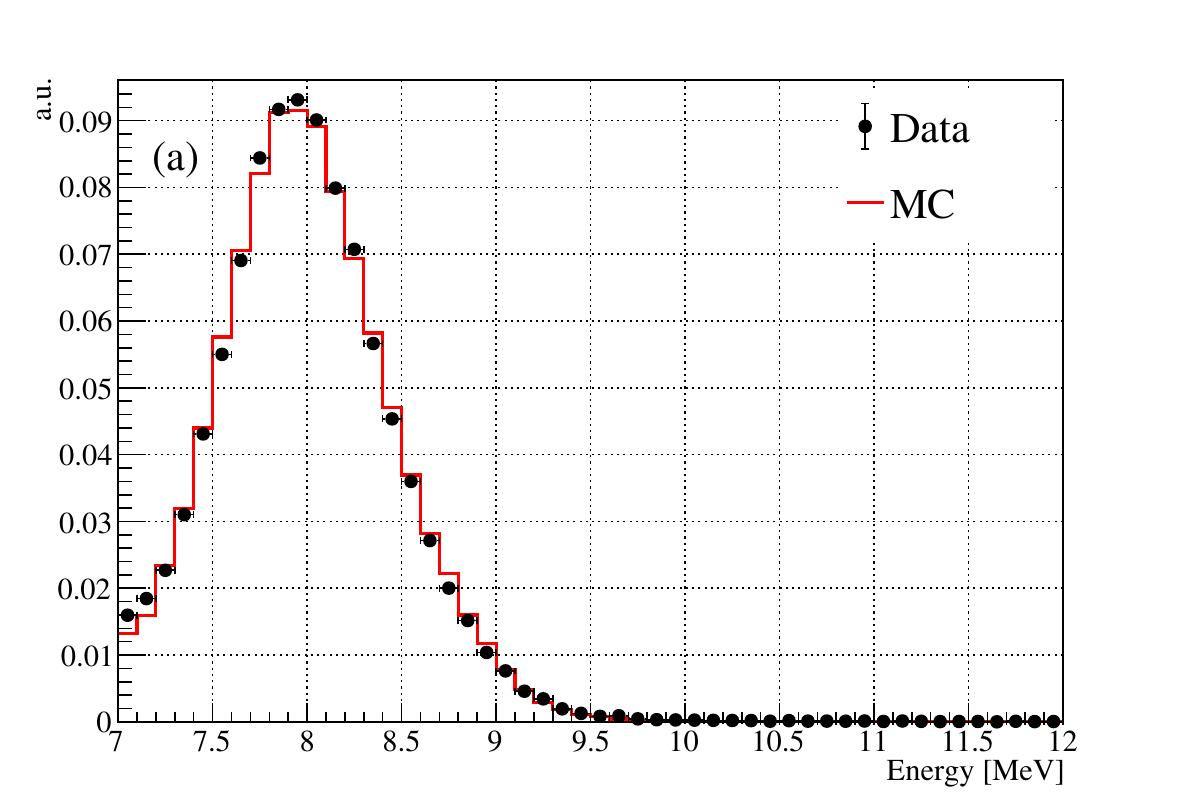}
			\qquad
	\end{minipage}
	\begin{minipage}[b]{0.45\textwidth}
       \includegraphics[width=0.96\textwidth]{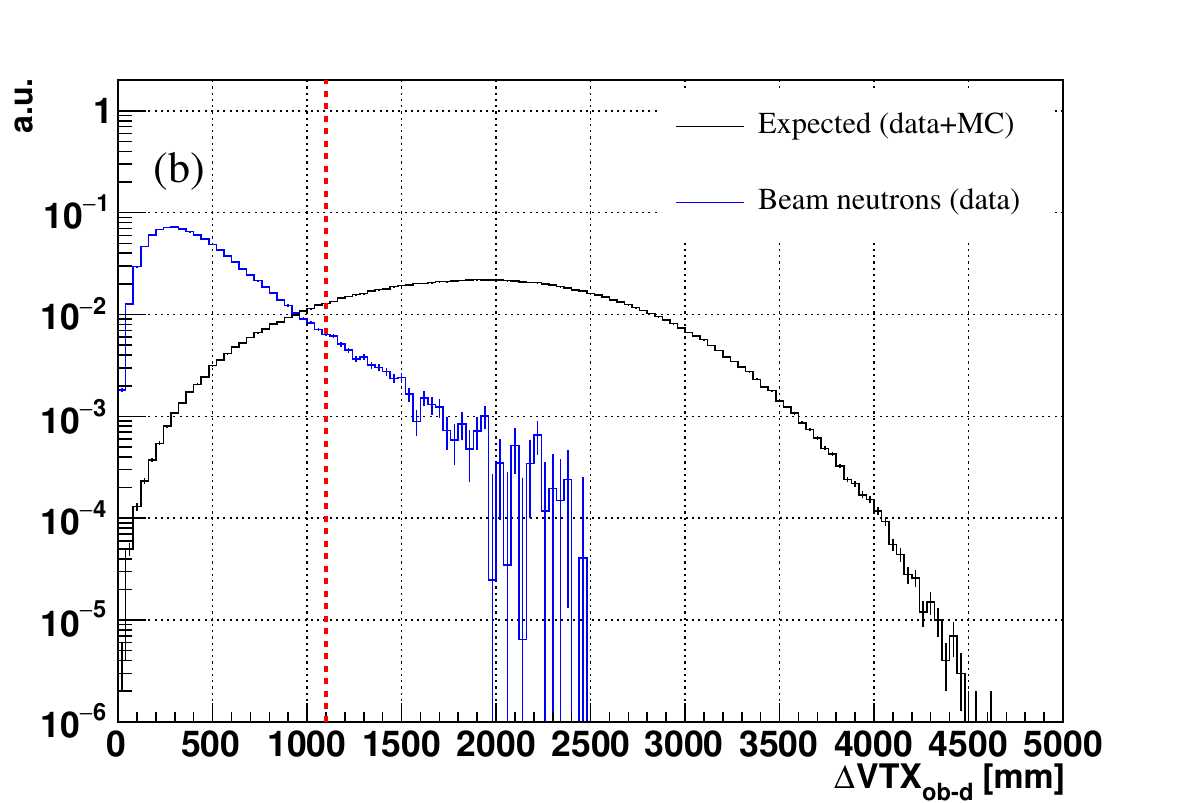}
			\qquad
    \end{minipage}       
	\caption{
    (a) Expected and observed reconstructed energy spectra of IBD delayed candidates.
    (b) Expected $\Delta$VTX$_{OB-d}$ distribution (black: data + MC) and that from 
    beam neutrons (blue: data). The red line 
indicates the applied selection threshold. All plots are area normalized.}
\label{Fig:Delayed}
\end{figure}

\indent
The delayed IBD candidates originate from neutron captures on Gd, 
releasing gamma rays with a total energy 
of approximately 8 MeV. Figure~\ref{Fig:Delayed}(a) shows the reconstructed energy spectra 
of simulated and observed n–Gd capture events, denoted as $E_{\mathrm{d}}$. The observed distribution is obtained 
from cosmogenic neutron samples, and good agreement is found between data and MC 
using this control sample. A delayed-energy selection of 
$7\le E_{\mathrm{d}}\le 12$~MeV is applied to identify IBD candidates.
The uncertainty in the delayed energy selection efficiency includes contributions from both the 
energy scale uncertainty and the n–Gd capture efficiency. Since the JSNS$^2$
 Gd–LS was donated by the Daya Bay experiment, the latter uncertainty value follows 
 that reported in Ref.~\cite{CITE:DayaBay}.

Beam-related neutron backgrounds can be suppressed using the prompt timing variable, 
$\Delta T_{beam-p}$, by selecting prompt candidates with higher energy due to  
the shorter time-of-flight from the mercury target to the detector. 
However, neutrons 
that are thermalized within the target region can produce delayed background 
candidates. To reject such events, JSNS$^2$ developed a spatial correlation rejection.  
If an activity that occurred during the beam time window ("$OB$" timing) is spatially 
close to an n–Gd capture vertex ("$d$" vertex), 
the event is identified as originating from a beam neutron. Thus, we required  
$\Delta$VTX$_{OB-d} \ge 110$~cm, where $\Delta$VTX$_{OB-d}$ is the spatial distance
between them.
Figure~\ref{Fig:Delayed}(b) shows the expected $\Delta$VTX$_{OB-d}$ distributions for 
IBD events (black, data+MC) and beam-neutron backgrounds (blue, data). The red line 
indicates the applied selection threshold. The IBD selection efficiency for this 
criterion is 94.1$\pm$0.1\%.
The event multiplicity during the beam time window 
is also taken into account; in cases with no beam-related activity, about half of beam spills,
no rejections on $\Delta$VTX$_{OB-d}$ are applied, resulting 
the signal selection efficiency is effectively 100\%.
The dominant systematic uncertainty arises from the vertex reconstruction accuracy, as discussed in 
Ref.~\cite{CITE:Cf}. The overall beam-neutron rejection efficiency achieved by this selection is 
approximately 93\%.

\subsection{IBD paired candidates}
\indent

\begin{figure}[htbp]
	\centering
	\begin{minipage}[b]{0.46\textwidth}
      \includegraphics[width=0.96\textwidth]{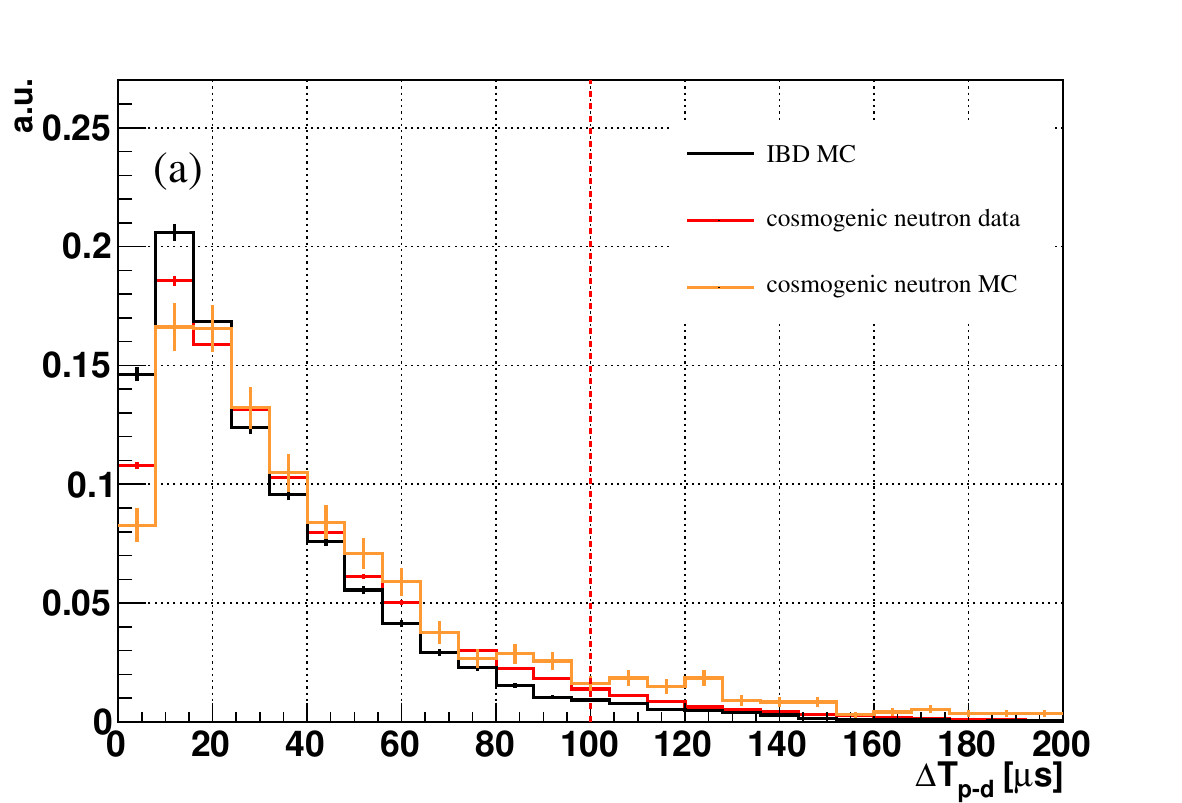}
			\qquad
	\end{minipage}
	\begin{minipage}[b]{0.46\textwidth}
       \includegraphics[width=0.96\textwidth]{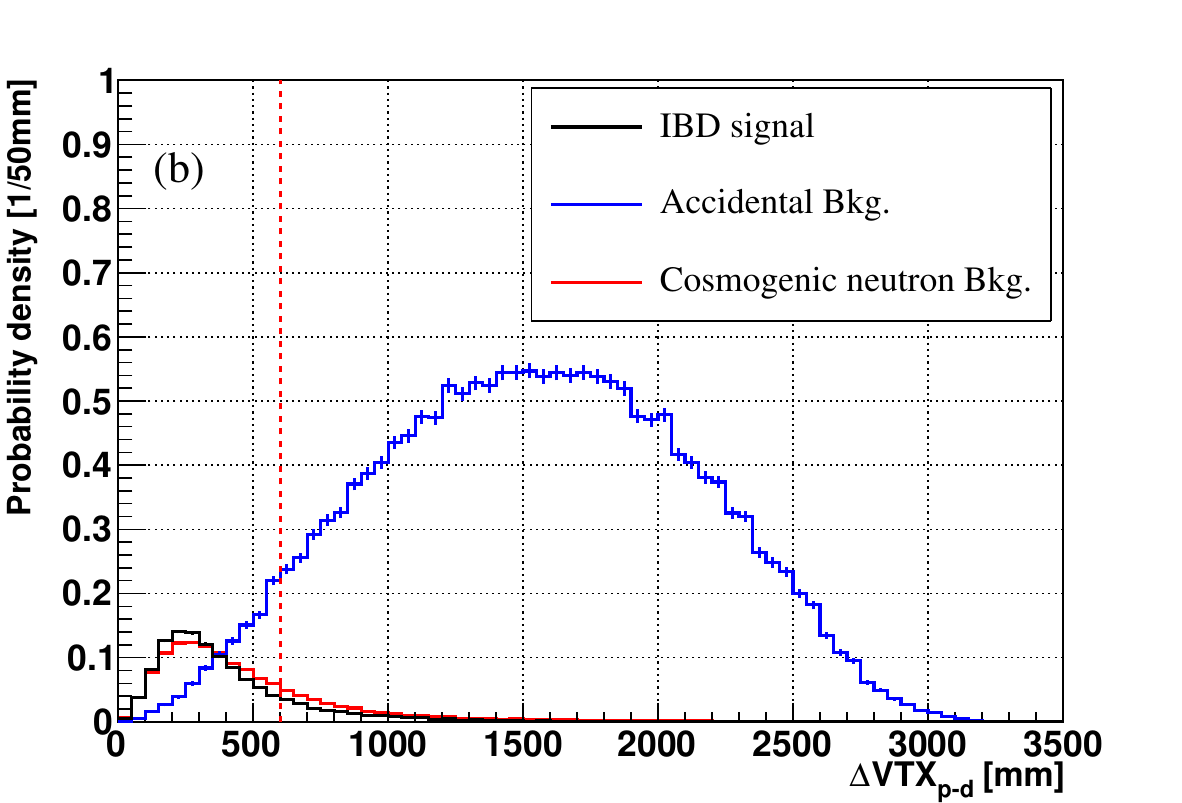}
			\qquad
    \end{minipage}       
	\caption{ (a) $\Delta T_{p-d}$ distributions for the IBD (MC: black), cosmogenic 
     neutron samples from data (red) and MC (orange). The plot is area normalized.
    (b) $\Delta$VTX$_{p-d}$. Black corresponds to the IBD signal,
    blue shows the accidental background and red shows the cosmogenic 
    neutrons. 
    For (b), all distributions are area normalized to one after applying the 
    $\Delta$VTX$_{p-d} < 60$~cm selection.
     }
\label{Fig:Paired}
\end{figure}

The timing and spatial correlations between the IBD prompt and delayed candidates provide powerful discrimination between signal and background events. 
Neutron captures on Gd in Gd-LS occurs with a mean time of
approximately 30 $\mu$s; therefore, events are required to satisfy $\Delta T_{p-d} < 100 ~\mu$s.
Figure~\ref{Fig:Paired} shows the $\Delta T_{p-d}$ distributions for IBD events (MC: black) 
and cosmogenic neutron samples obtained from data (red) and MC (orange).
Since no IBD data sample is available to directly assess the systematic uncertainty, the difference 
between data and MC for the cosmogenic neutron sample is used to estimate it, assuming similar 
behavior between neutron and IBD events. The efficiency derived from the IBD MC is 
96.6$^{+3.4}_{-5.6}$\%.

Applying the same approach to the spatial correlation selection, $\Delta$VTX$_{p-d}$, yields an 
efficiency of 83.4$\pm$10.0\%. 
The distributions for the IBD MC, the cosmogenic neutron background 
data and the accidental background are shown in Fig.~\ref{Fig:Paired}(b). 

Signal and background discrimination is further enhanced by a likelihood ratio constructed from 
six variables: $\Delta T_{beam-p}$, $E_{\mathrm{d}}$, $\Delta T_{p-d}$, $\Delta$VTX$_{p-d}$, 
and the reconstructed vertex coordinates y (beam direction) and 
z (vertical direction). This multivariate likelihood selection depends sensitively on 
the background modeling, which is discussed in detail in a later section.

\subsection{Rejection of cosmic rays}
\indent
JSNS$^2$ is an above-ground experiment, with its detector located on the third floor of the MLF 
building. Consequently, suppression of cosmogenic backgrounds is essential. 
The detector’s veto region and additional techniques are fully utilized for this purpose.

For the muon veto, the summed charge of the twelve PMTs in the top veto region is required 
to be less than 30 photoelectrons (p.e.), and that of the twelve PMTs in the bottom region 
less than 40 p.e., within a 200 ns time window relative to the activity in the target volume. 
This requirement is applied to both prompt and delayed candidates. 
The inefficiency of this veto for IBD signals is evaluated using cosmogenic control samples: 
Michel-electrons for prompt candidates and gammas from n-Gd captures for delayed candidates. 
The resulting IBD selection efficiency from this veto is 92.8$\pm$0.5\%.
After the muon veto, cosmogenic Michel-electrons remain as a dominant background, 
particularly for prompt candidates. To reject them, parent muons are searched for 
within 10 $\mu$s before the target activity (“Michel-e veto”). The criterion for identifying 
a parent muon is a summed charge exceeding 100 p.e. in either the top or bottom twelve PMTs. 
With a muon rate of approximately 2 kHz, the accidental-coincidence probability between a muon 
and an IBD event occurring within the 10 $\mu$s 
search window in the IBD delayed region is about 2\%. 
Considering a trigger dead time for the IBD prompt candidates, the inefficiency of the Michel- 
electron veto is about 0.9\%, corresponding to a signal selection efficiency of 97.0$\pm$0.03\%
in total.

Due to the trigger dead time, 
some Michel-electrons still contaminate the IBD prompt candidate sample. 
To remove these, JSNS$^2$ developed an innovative method to tag muons and Michel-electrons using 
the FADC baseline~\cite{CITE:baseline}. High-energy particles cause a measurable shift in the 
FADC baseline following their passage. This feature allows us to tag Michel-electrons with their
hidden parent muons with 99.8\% efficiency. 
However, this method introduces a small probability for real IBD events to be incorrectly 
rejected due to a large charge event 
preceding it. High-energy beam neutrons during the beam window can also induce
baseline shifts, resulting in the additional inefficiency. Accounting for these effects, the IBD 
selection efficiency on this baseline cut is 86.1$\pm$1.4\%.

The central chimney structure of the detector provides necessary access to the inner 
volume, but lacks veto coverage.
If a muon traverses this region and produces an event within the 20–60 MeV energy range, 
neither the veto nor the FADC baseline method can identify it. Because scintillation light is 
hidden by the chimney structure, 
the reconstructed vertex shows poor fit quality~\cite{CITE:JSNS2PSD}. 
This “goodness” parameter is used to reject such background events. The corresponding IBD 
selection efficiency with this criterion is 98.7$\pm$0.1\%.

The dominant systematic uncertainty for the cosmic-ray veto arises from detector stability. 
The uncertainty for the Michel-electron rejection using the FADC-baseline shift includes the difference observed between a sideband-region sample (described later) and a single-triggered control sample.

\section{Backgrounds}
\indent
Two main background sources are considered in this analysis: 
accidental coincidences and cosmogenic neutrons. 

The accidental background is described in detail in Ref.~\cite{CITE:accidental}. 
The rate of accidentals backgrounds passing the IBD prompt selection in 
Table~\ref{tab:EventSelection} is $[$(8.5$\pm0.1)\times 10^{-5}$/beam spill$]$,
which is dominated by cosmogenic $\gamma$-rays. The corresponding rate for accidentals passing 
the IBD delayed selection is $[$(1.9$\pm 0.01) \times 10^{-2}$/beam spill$]$. The primary
contributions to this background are (a) residual beam neutrons ($\sim$50\%), 
(b) beam-induced $\gamma$ rays ($\sim$25\%), and (c) cosmogenic $\gamma$-rays.
To suppress accidental coincidences, a spatial correlation requirement of 
$\Delta$VTX$_{p-d} < 60$~cm is applied,
which plays a crucial role in reducing uncorrelated events. 
Using a dedicated calibration run with beam-scheduled timing triggers and the spill-shift 
method~\cite{CITE:accidental}, the rejection factor for accidental backgrounds is found to be 
approximately 94.7\% (Fig.~\ref{Fig:Paired}(b)).

Neutrons constitute a correlated background, fully mimicking the IBD signature and therefore not 
removed by the standard IBD selections. Their rejection relies primarily on (a) PSD and (b) a
likelihood-ratio discrimination method. The cosmogenic neutron control sample is 
defined by events occurring 
later than 1 ms after the beam start. PSD application reduces this background by a 
factor of $\sim$300, and the likelihood-ratio provides an additional suppression by a factor of a few as described later.

The event rates and variable distributions for both accidental and correlated backgrounds are 
evaluated using in situ, data-driven control samples, not MC simulation. These samples include the 
correct background rejection factors a priori, and minimize systematic uncertainties 
associated with modeling, energy calibration, and time-dependent effects. 

In addition to the backgrounds discussed above, a small contribution arises from neutrino-induced processes. The CNgs reaction produces background events when the subsequent $\beta$
decay of $^{12}$N$_{g.s.}$ occurs within 100 $\mu$s. The in-situ measurements reported in Ref.~\cite{CITE:JSNS2CNgs} enable a data-driven estimation of this background.
The beam intrinsic $\bar{\nu}_{e}$ component from the $\pi^{-} - \mu^{-}$ decay chain 
at the mercury target is evaluated solely by MC simulation, since the $\pi^{-}$ production 
rate from 3 GeV proton interactions with mercury has not been directly measured. 
The uncertainty on this background is conservatively assigned as 50\%, 
derived from the discrepancy between the measured and simulated CNgs rate from the 
$\pi^{+} - \mu^{+}$ chain.
Since these neutrino-induced backgrounds become relevant only after suppression of the dominant 
cosmogenic-neutron and accidental backgrounds, the corresponding rejection methods 
applied after the selections listed in Table~\ref{tab:EventSelection} for the 
dominant ones are described in this section.

\subsection{Multivariate Likelihood-ratio Selection}
\indent
After applying the event selection summarized in Table~\ref{tab:EventSelection}, several variables
exhibit distinct differences between the IBD signal and background samples. 
Although these variables are not directly used for binary selection cuts, they provide 
significant discriminating power for the multivariate likelihood-ratio analysis.

Six input variables are used to construct the likelihood ratio: 
(a) $\Delta T_{beam-p}$, (b) $\Delta T_{p-d}$, (c) $\Delta$VTX$_{p-d}$, 
(d) $E_{\mathrm{d}}$, and 
the reconstructed vertices on (e) y (beam direction) and (f) z (vertical) coordinates.   
Cosmogenic backgrounds yield flat timing distributions in (a) and (b),  
but are concentrated at higher z, as expected for cosmic rays. In contrast, IBD events follow 
exponential timing and a flat z distribution. The accidental background shows no spatial 
correlation between prompt and delayed candidates, resulting in the uniform phase-space 
pattern visible in (c).
For variables (b), (c), (e), (f), 
MC simulations are used to model the IBD signal, while for 
(a), corrected timing information from another neutrino source~\cite{CITE:JSNS2KDAR} is 
employed. Discrimination between accidental and IBD events is primarily achieved using all 
six variables, whereas separation between neutron-induced and IBD events relies mainly on 
(a) and (e).
Probability density functions (PDFs) for each component are constructed from these six 
variables. To ensure statistical independence, half of the background control sample is 
used to build the PDFs, while the remaining half is reserved for evaluating the 
likelihood-ratio scores.

A two-dimensional likelihood-ratio function is constructed to separate the signal 
from backgrounds: $LLK_{signal}~ - LLK_{accidental}$ and $LLK_{signal}~ - LLK_{neutron}$.
$LLK$ denotes the negative log-likelihood, and subscripts indicate the type of sample. 
In this analysis, the signal region of the two-dimensional log-likelihood ratio is defined as the 
area where both log-likelihood scores satisfy $\le 0$.

The likelihood separation efficiency for IBD events is obtained as
70.5$\pm$1.5\%. This method suppresses approximately 70\% of neutron-induced backgrounds and 
84\% of accidental backgrounds. The dominant systematic uncertainty arises from the modeling 
of the IBD MC.

\subsection{Energy Side-band Study}
\indent
To validate the analysis methodology prior to extracting the search results, we performed an energy side-band study. 
Relative to the signal region defined in Table~\ref{tab:EventSelection}, which applies
$20\le E_{\mathrm{p}} \le 60$~MeV and $7\le E_{\mathrm{d}}\le 12$~MeV, 
four independent side-band regions were defined as follows: (A) $E_{du}$ : delayed 
energy 12–20 MeV, (B) $E_{dl}$ : 5–7 MeV, (C) $E_{pu}$ : prompt energy 60–100 MeV, and (D) $E_{pl}$ 
: 12–20 MeV, where the subscripts $"u"$ and $"l"$ denote "upper" and "lower" energy 
regions, respectively.
All other selection criteria listed in Table~\ref{tab:EventSelection} 
were applied unchanged, except for the $E_{pl}$
region, where the lower bound of $\Delta T_{beam-p}$
was increased from 2 to 3 $\mu$s to suppress beam-neutron contamination.

In these side-band regions, the expected selection efficiency for IBD events associated with the LSND anomaly, as well as for other neutrino-induced backgrounds, is at least an order of magnitude smaller than in the nominal selection. These contributions are therefore assumed to be negligible in the predicted event counts.

The number of selected events and the distributions of relevant variables were compared between data 
and prediction in four cases: (1) without pulse-shape discrimination (PSD) or log-likelihood ratio (LLK), (2) 
with LLK only, (3) with PSD only, and (4) with both PSD and LLK applied. The PSD threshold was set to 
reject more than 99.5\% of neutron backgrounds, and the LLK criterion required both likelihood 
scores to be less than zero (i.e., signal-like events). Table~\ref{tab:sideband} summarizes 
the observed and predicted event counts in each side-band region and for each case. 
Predictions include systematic uncertainties but not statistical (Poisson) 
uncertainties. For the $E_{du}$
region, only case (1) is reported due to the limited number of events.

The predictions, composed of accidental and cosmogenic neutron background components 
estimated from 
data-driven control samples, are in good agreement with the observations across all regions 
and selection stages. The shapes of the observed and predicted distributions for all variables 
also show a consistent behavior, validating the analysis procedure before applying it to the 
signal region.
\begin{table*}[htb]
    \centering
    \caption{\label{tab:sideband}
     Comparison of the number of events between the observations and the predictions for each 
     side-band region and each analysis step.
     The predictions include only systematic uncertainties.}
    \vspace{3pt}
    \begin{tabular}{|c|cc|cc|cc|cc|}\hline
        cases & $E_{du}$ obs. & pred. & $E_{dl}$ obs. & pred. & $E_{pu}$ obs. & pred. & $E_{pl}$ obs. & pred.  \\ \hline
        w/o PSD w/o LLK & 3 & 3.5$\pm$0.7 & 91 & 106.3$\pm$2.7 & 703 & 711.7$\pm$16 & 191 & 184.6$\pm$5.2 \\ 
        w/o PSD w/ LLK & -- & -- & 25 & 27.4$\pm$0.8 & 240 & 213.6$\pm$4.8 & 49 & 50$\pm$1.5 \\
        w/ PSD w/o LLK & -- & -- & 2 & 5.1$\pm$0.3 & 5 & 5.4$\pm$0.3 & 6 & 3.3$\pm$0.2 \\
        w/ PSD w/ LLK  & -- & -- & 0 & 0.9$\pm$0.1 & 1 & 1.2$\pm$0.2 & 0 & 0.5$\pm$0.08 \\ \hline
    \end{tabular}
\end{table*}

\section{Expected number of signal events}
The expected number of events associated with the LSND anomaly in JSNS$^2$ is 
described in this section. Two different estimation methods are described:
a model-independent method based on LSND’s anomaly, 
and a method that assumes the excess is from short-baseline neutrino oscillations.

\subsection{Normalization-based calculation}
\indent

The normalization-based prediction employs Eq.~\ref{Eq:LSND},
\begin{equation}
    N_{JSNS^2} = \frac{\Phi_{JSNS^2} \cdot \sigma \cdot N_{T~JSNS^2} \cdot \epsilon_{JSNS^2}}{\Phi_{LSND} \cdot \sigma \cdot N_{T~LSND} \cdot \epsilon_{LSND}} \times N_{LSND},
\label{Eq:LSND}
\end{equation}
where $\Phi$ denotes the time-integrated neutrino flux at the detector in units of neutrinos 
per cm$^2$, 
$\sigma$ is the IBD cross section, $N_T$ is the number of hydrogen 
atoms serving as antineutrino targets, $\epsilon$ is the event selection efficiency, and 
$N$ is the number of events. We assume that the energy spectrum of $\bar{\nu}_e$
is identical for LSND and JSNS$^2$ in this calculation. 
Under this assumption, the neutrino flux scales with 
the inverse square of the baseline, while the IBD cross section is common to both experiments
and cancels out.

Table~\ref{tab:LSNDsignal} summarizes the parameters used in Eq.~\ref{Eq:LSND} for 
LSND~\cite{CITE:LSND} and JSNS$^2$. 
\begin{table}[htb]
    \centering
    \caption{\label{tab:LSNDsignal}
     A comparison between LSND, KARMEN and JSNS$^2$ on 
     a neutrino flux (number of neutrinos per cm$^2$), a number of targets for anti 
     electron neutrinos ($N_{T}$), the selection efficiency (\%). The observed number
    of events in LSND, and the expected number of events for JSNS$^2$ and KARMEN from
    the LSND anomaly are also shown ($N_{eve}$).
    }
    \vspace{3pt}
    \begin{tabular}{cccc}\hline
        item & LSND & KARMEN & JSNS$^2$ \\ \hline
        $\nu$ flux ($\#\nu/cm^2$) & 1.26$\times 10^{14}$ & 6.89$\times 10^{13}$ & 5.47$\times 10^{13}$ \\ 
        (POT & 1.81$\times 10^{23}$ & 6.05$\times 10^{22}$ & 0.82$\times 10^{22}$ ) \\
        ($\# \mu$/p & 0.079 & 0.0448$\pm$0.0030 & 0.48$\pm$0.17 ) \\
        (baseline & 30 m & 17.7 m & 24 m ) \\
                    &   &    &              \\
        $N_T$  & 7.4$\times 10^{30}$  & 4.5$\times 10^{30}$ & 7.48$\times 10^{29}$   \\
                    &   &    &              \\        
        efficiency (\%)  & 42.0$\pm$3.0 & 19.1$\pm 1.45$ & 12.5$^{+2.1}_{-2.2}$ \\ \hline
        $N_{eve}$ & 87.9$\pm$23.1 &  12.6$\pm$3.3 & 1.1$\pm$0.5\\ \hline
    \end{tabular}
\end{table}
The quantity $\# \mu$/p represents the number of positive muons 
produced per incident proton at the neutrino production target, and $N_{eve}$
is the observed number of events in LSND, and the expected number of events from 
the LSND anomaly with this normalization for JSNS$^2$. 
In the present analysis, 
JSNS$^2$ benefits from a higher antineutrino flux per proton due to the higher 
proton beam energy, whereas LSND achieved a larger effective target mass and 
higher detection efficiency.

The excess of $\bar{\nu}_e$ events observed by LSND is 87.9$\pm$22.4 (stat)$\pm$6.0 (syst). 
Using Eq.~\ref{Eq:LSND}, this corresponds to an expected number of 1.1$\pm$0.5
LSND anomaly events in JSNS$^2$. The uncertainty is dominated by the muon production 
measurement in JSNS$^2$, which is limited by statistics and fiducial volume definition, 
as well as by the uncertainty of the LSND measurement itself.

For reference, Table III also includes the corresponding parameters for the KARMEN 
experiment~\cite{CITE:KARMEN2}, which also directly tested the LSND anomaly. 
KARMEN observed 15 events, compared with an expectation of $28.4 \pm 6.3$ events 
under the LSND-anomaly hypothesis (including backgrounds), corresponding to a
$2.1\sigma$  tension.
In this context, JSNS$^2$ plays an essential role
in clarifying the tension between the LSND and KARMEN results.

\subsection{Oscillation-based calculation}
\indent

Beyond the normalization-based discussion above, neutrino oscillation effects at short 
baselines can also be considered separately. 
In this case, 
the oscillated $\bar{\nu}_{e}$ originates from $\bar{\nu}_{\mu}$ produced in the reaction 
$\mu^{+} \to e^{+} + \nu_{e} + \bar{\nu}_{\mu}$ in the mercury target. The energy
spectrum ranges from 0 to 52.8 MeV, 
and is weighted by an oscillation probability.
We assume two-flavor neutrino oscillations between
$\bar{\nu}_{\mu}$ and $\bar{\nu}_{e}$, expressed as
\begin{equation}
    P(\bar{\nu}_{\mu} \to \bar{\nu}_{e}) = \sin^2(2\theta_{\mu e}) \cdot \sin^2 \left( \frac{1.27\Delta m^2 L}{E_{\bar{\nu}}}\right),
\end{equation}
where $P$ is the oscillation probability, $\theta_{\mu e}$ is the mixing angle between 
muon and electron antineutrinos, $\Delta m^2$ (eV$^2$) is the squared mass difference 
between the two mass eigenstates, $E_{\bar{\nu}}$ is the antineutrino energy (MeV), and $L$
is the distance between the neutrino production and detection points (meter).

The IBD cross section is known with high precision (e.g.: \cite{CITE:IBD}). 
The IBD reaction produces a positron with a kinetic energy of 
($E_{\bar{\nu}_{e}} -$ 1.8) MeV approximately, 
which constitutes the prompt signal. Including the contribution 
from positron annihilation, the visible (deposited) energy in the 
detector is approximately
$E_{\mathrm{vis}} \simeq T_{e^+} + 2 m_e$, 
and the reconstructed prompt energy $E_p$ is derived from this 
visible energy.

Assuming the LSND best-fit oscillation parameters, 
(sin$^2 2\theta$, $\Delta m^2$) = (0.003, 1.2), reported in 
Ref.~\cite{CITE:LSND}, the expected number of events in JSNS$^2$ is 1.2$\pm$0.4.

\section{Results}
\indent
Table~\ref{tab:signal} summarizes the observed and predicted numbers of events in the IBD 
signal region defined in Table~\ref{tab:EventSelection} for each analysis case, which has the same definition as those in side-bands. 
The predictions assume no LSND anomaly contribution, while including the intrinsic 
$\bar{\nu}_{e}$ and the CNgs backgrounds. 
A total of 2 events are observed, consistent with the background-only expectation of 
2.3$\pm$0.4 events. No clear excess of $\bar{\nu}_{e}$ events is seen with the present 
statistics. The expected number of $\bar{\nu}_{e}$ events due to LSND anomaly defined
by the normalization-base calculation is 1.1$\pm$0.5 as mentioned above.

After applying both PSD and LLK selections, the predicted background composition consists of cosmogenic neutrons (0.6$\pm$0.2 events), accidental coincidences (1.00$\pm$0.01), 
CNgs (0.08$\pm$0.02) and intrinsic beam-related backgrounds (0.6$\pm$0.3).  

\begin{figure}[htbp]
	\centering
	\begin{minipage}[b]{0.48\textwidth}
      \includegraphics[width=0.96\textwidth]{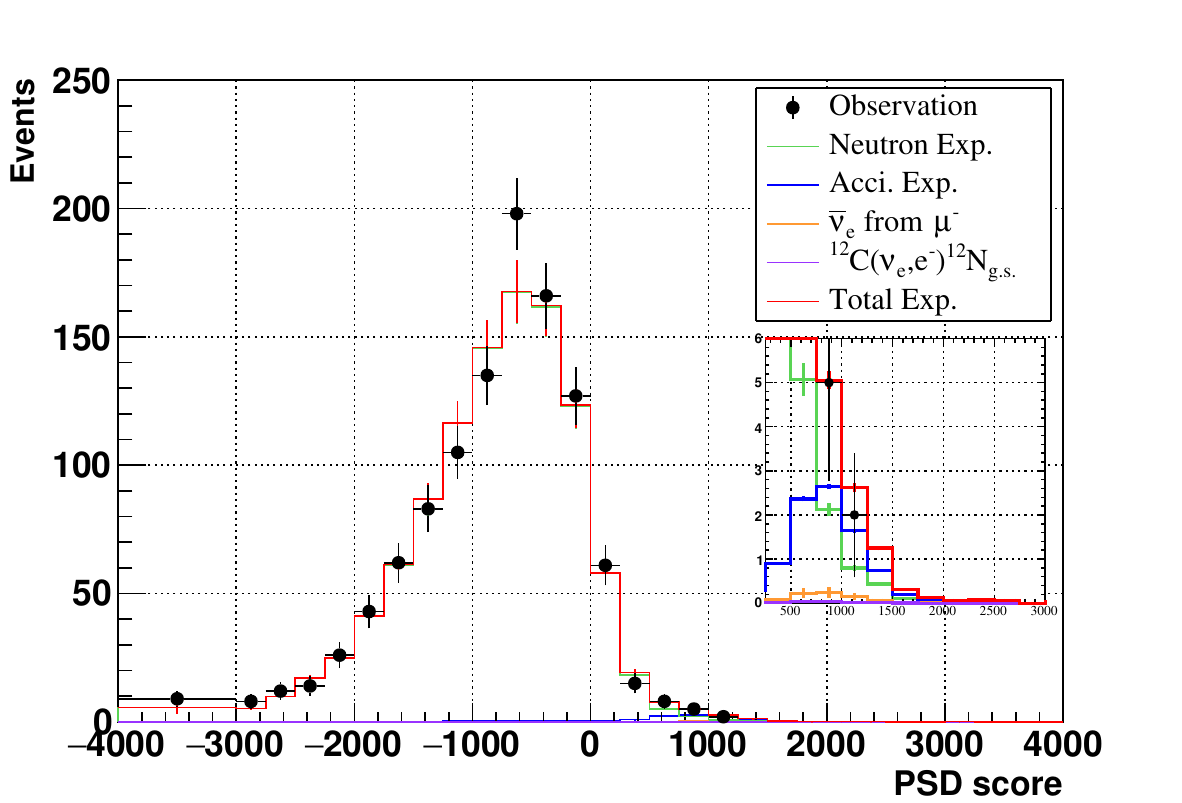}
			\qquad
	\end{minipage}
	\caption{
    PSD score distribution after applying the selection criteria listed in 
    Table~\ref{tab:EventSelection}. 
    The inset shows an expanded view of the same distribution. 
    The vertical axis indicates the number of events.
    }
\label{Fig:PSDscore}
\end{figure}

\begin{figure}[htb!]
	\centering
    \begin{minipage}[b]{0.43\textwidth}
       \includegraphics[width=0.93\textwidth]{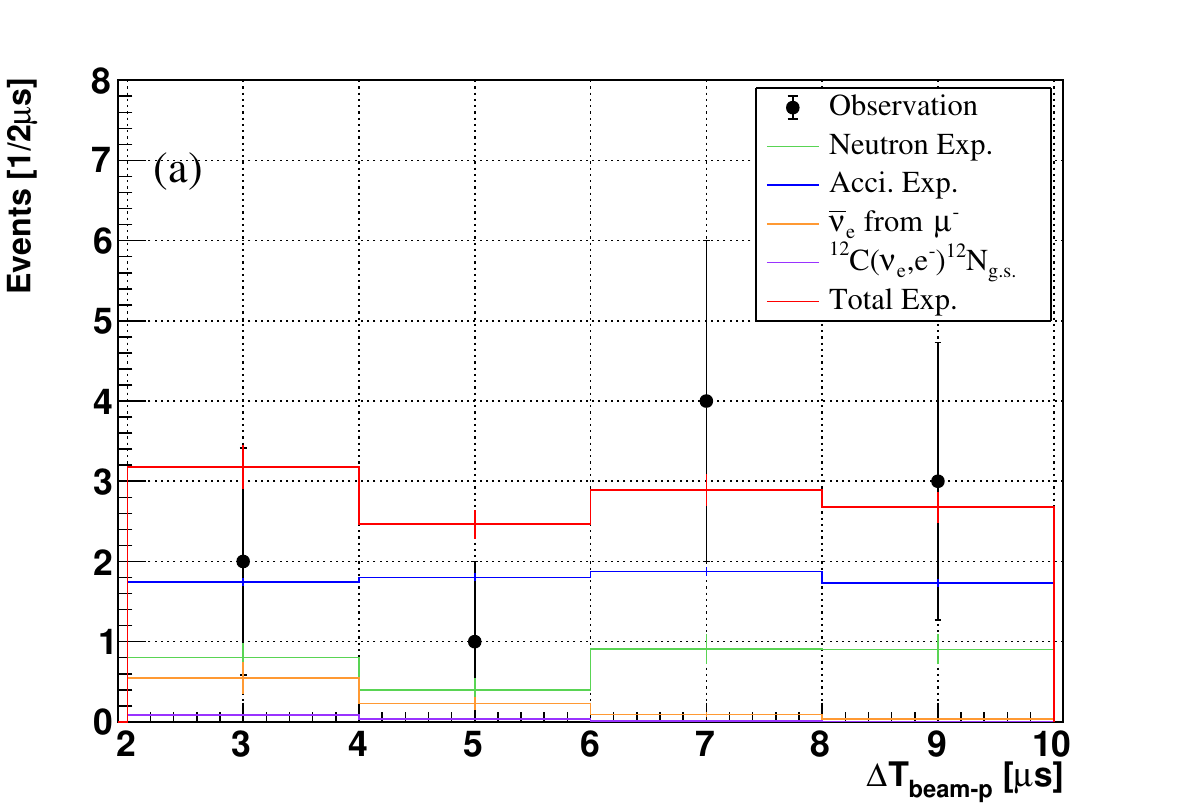}
			\qquad
    \end{minipage} 
	\begin{minipage}[b]{0.43\textwidth}
       \includegraphics[width=0.93\textwidth]{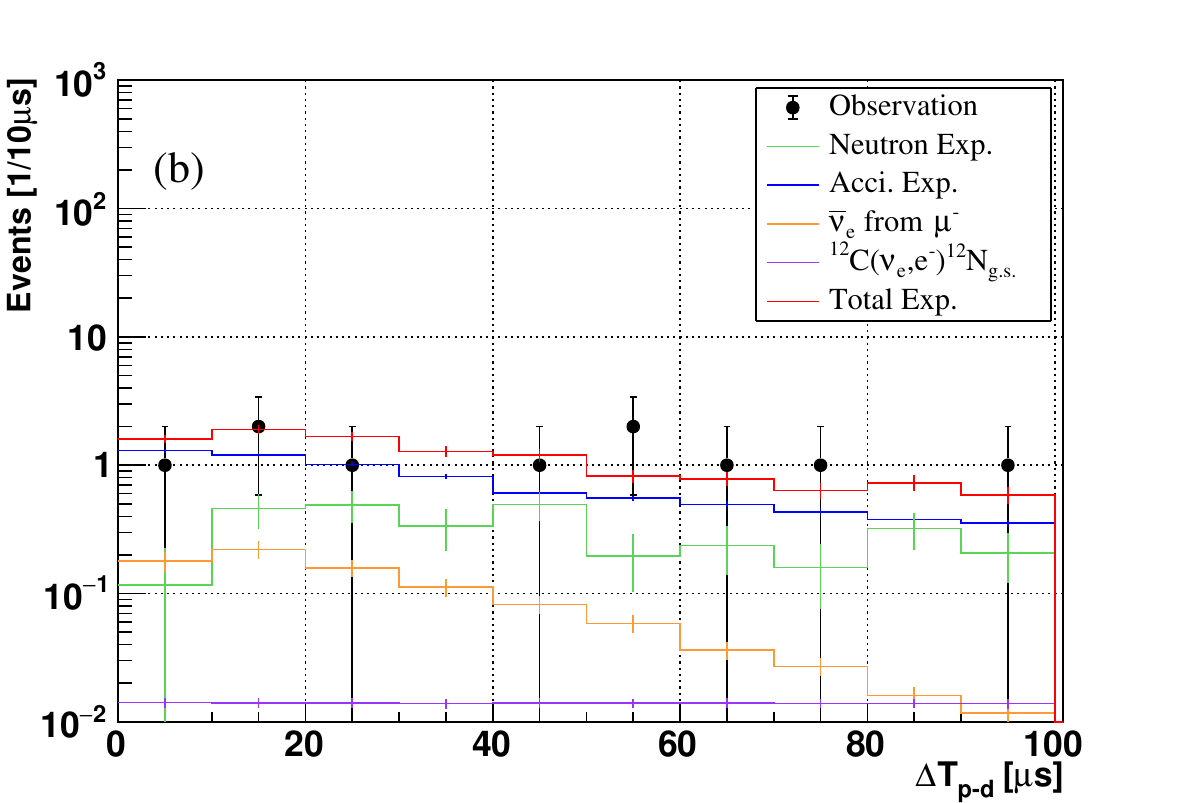}
			\qquad
    \end{minipage}    
	\begin{minipage}[b]{0.43\textwidth}
      \includegraphics[width=0.93\textwidth]{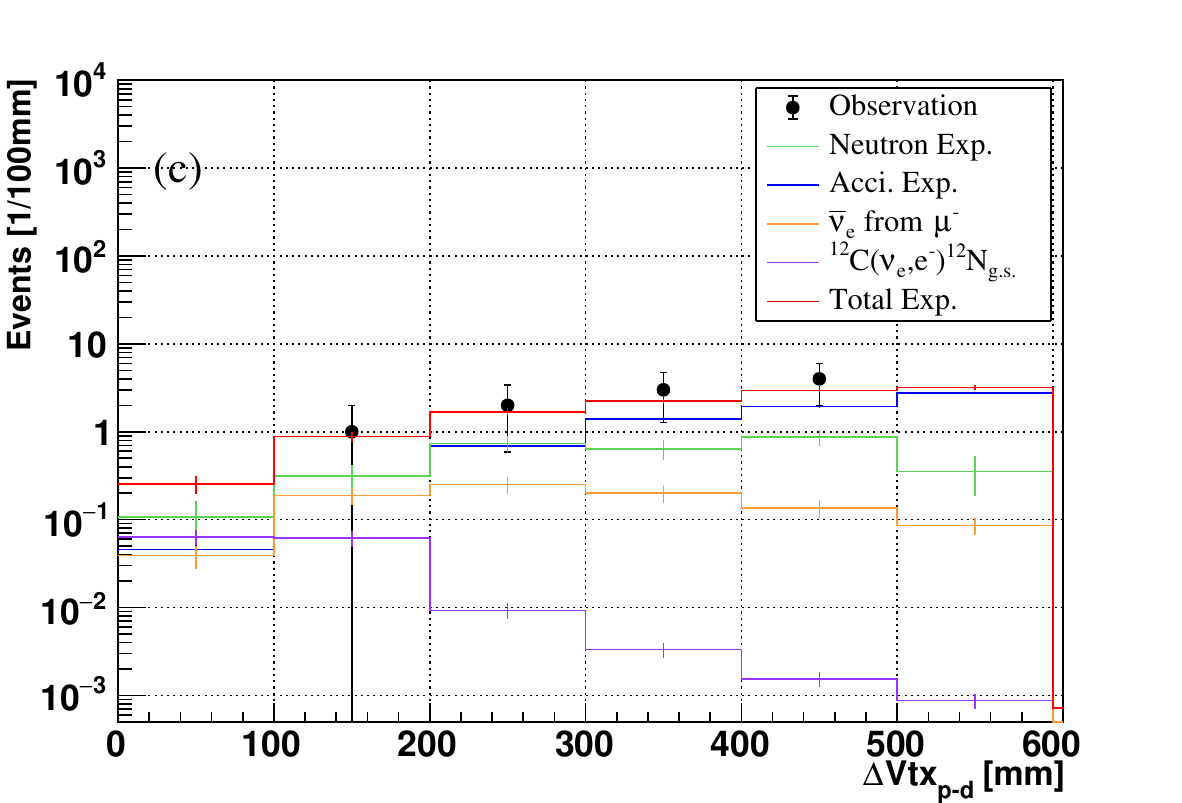}
			\qquad
	\end{minipage}
    \begin{minipage}[b]{0.43\textwidth}
       \includegraphics[width=0.93\textwidth]{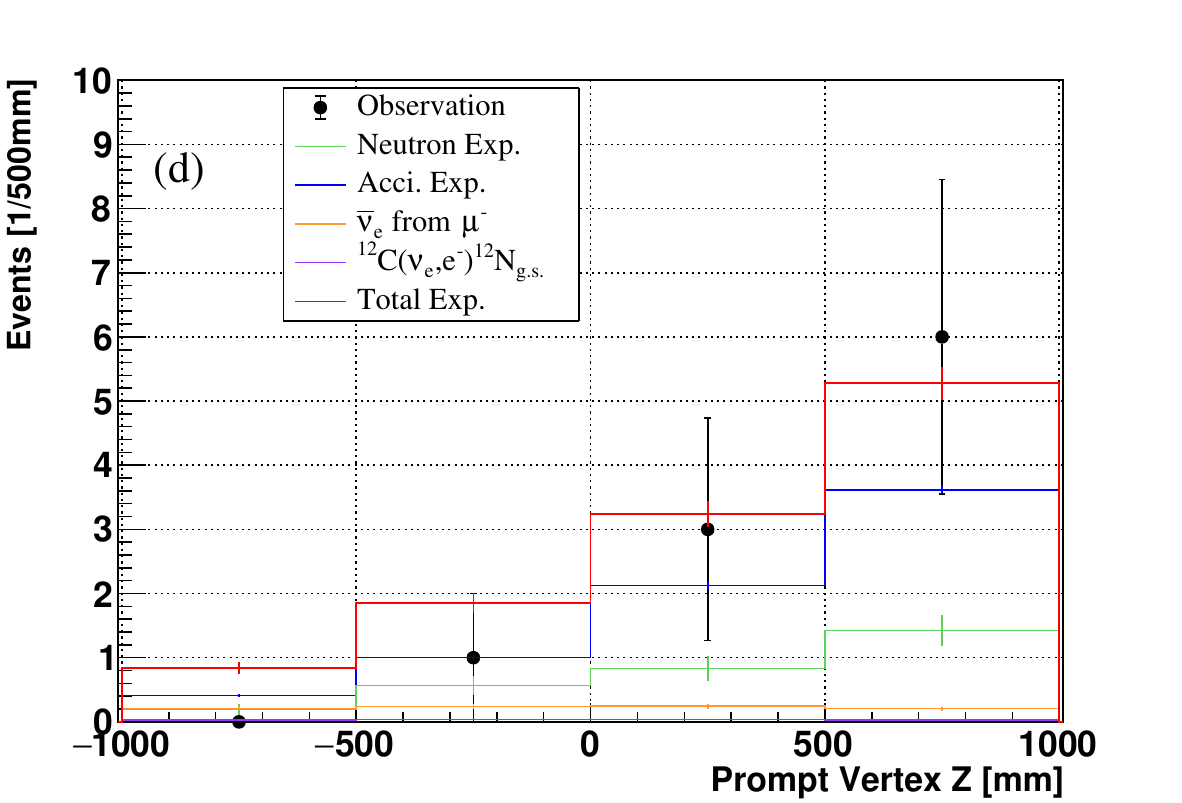}
			\qquad
    \end{minipage}       
	\caption{Distributions of (a) $\Delta T_{beam-p}$, (b) $\Delta T_{p-d}$, 
    (c) $\Delta$VTX$_{p-d}$ and (d) z vertex (height direction) after the PSD application. 
    The vertical axes indicate number of events.}
\label{Fig:results}
\end{figure}

\begin{figure}[htbp]
	\centering
	\begin{minipage}[b]{0.46\textwidth}
      \includegraphics[width=0.96\textwidth]{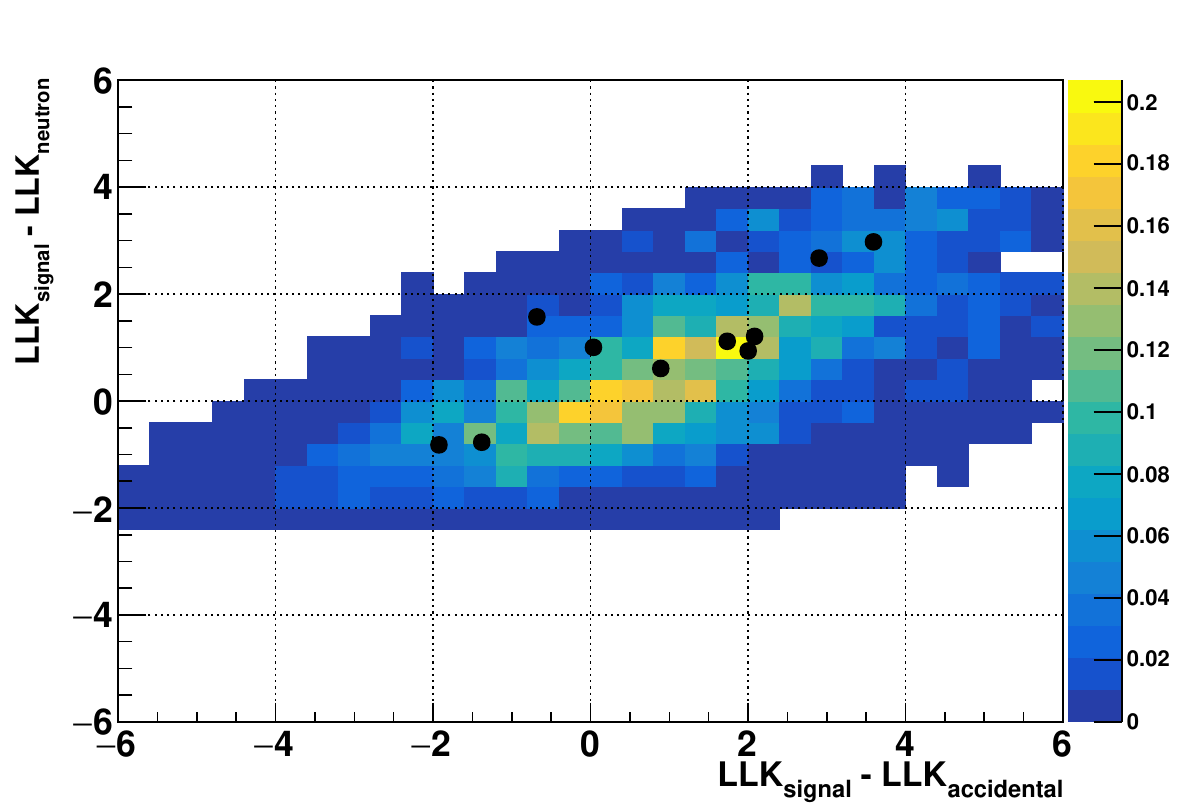}
			\qquad
	\end{minipage}
	\caption{
    Two-dimensional log-likelihood ratio distribution for the observation (black points) and the expected background distribution, consisting of backgrounds, shown with the predicted number of events in main text.
     }
\label{Fig:LLK}
\end{figure}

Figure~\ref{Fig:PSDscore} shows the PSD score distribution after applying all selection criteria 
except for the PSD cut, while Fig.~\ref{Fig:results} presents the distributions of 
reconstructed 
timing and vertex after the PSD selection. The applied PSD score 
threshold ranges from 550 to 730 depending on energy. The observed distributions 
agree well with the expected ones across all variables, confirming the 
consistency of the event selection and background modeling.
Figure~\ref{Fig:LLK} gives the 2D LLK distributions for the observations 
overlaid with the backgrounds, which have a good agreement. 
The LLK selection requires both the horizontal and vertical log-likelihood scores 
to be less than zero, corresponding to the signal-like region. 

\begin{table}[htb]
    \centering
    \caption{\label{tab:signal}
    Comparison of the number of events between the observation and the prediction for the signal-region in each analysis step.
    The predictions include only systematic uncertainties.}
    \vspace{3pt}
    \begin{tabular}{ccc}\hline
        cases & observation & prediction \\ \hline
        w/o PSD w/o LLK  & 1079 & 1063.0$\pm$25.7 \\ 
        w/o PSD w/ LLK  & 304  & 315.8$\pm$8.0   \\
        w/ PSD w/o LLK  & 10   & 11.2$\pm$0.7 \\
        w/ PSD w/ LLK   & 2    & 2.3$\pm$0.4 \\ \hline
    \end{tabular}
\end{table}


\subsection{Neutrino oscillation interpretation}
\begin{figure}[htb!]
	\centering
	\begin{minipage}[b]{0.48\textwidth}
      \includegraphics[width=0.97\textwidth]{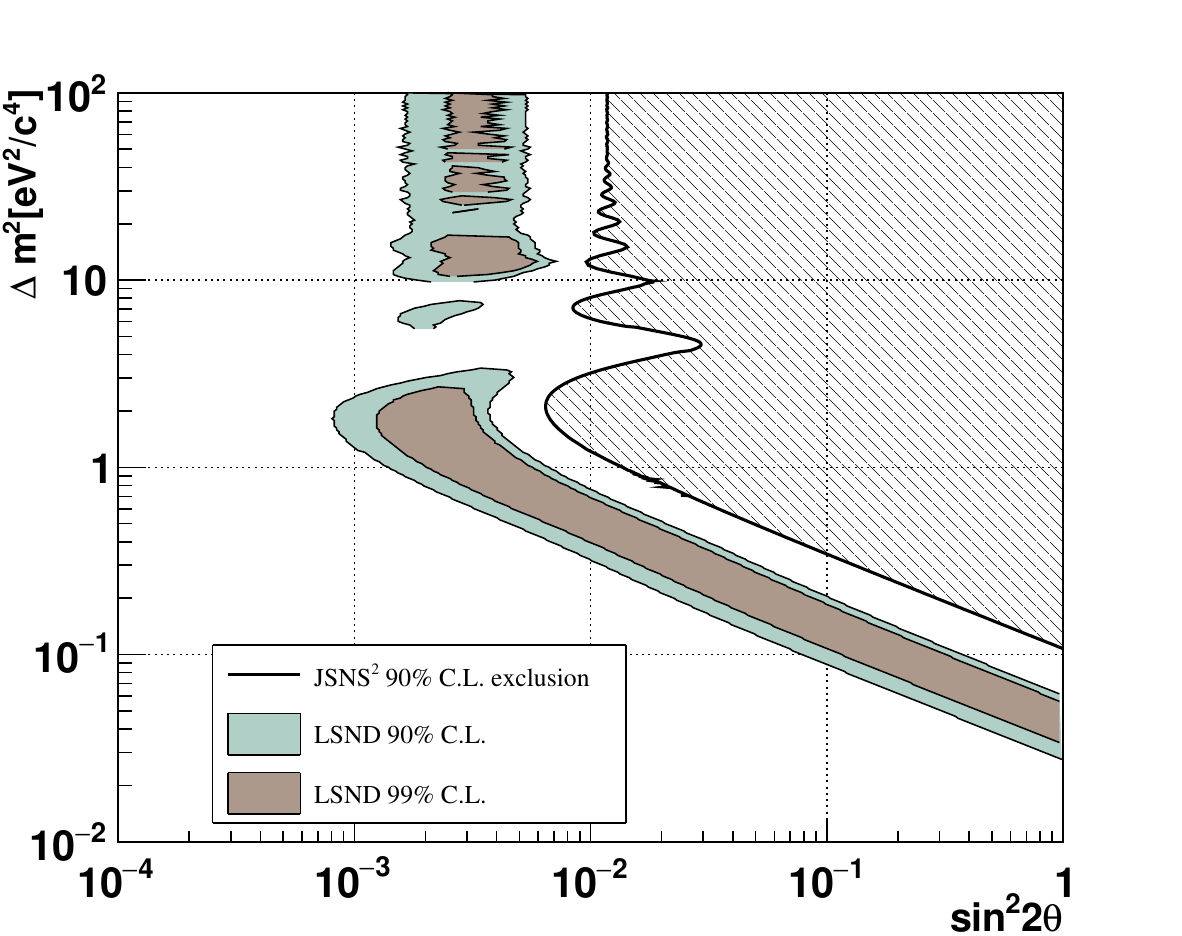}
			\qquad
	\end{minipage}     
	\caption{
    90\% C.L. exclusion limit derived from the observation (black line). 
    Shaded region shows the excluded region of the neutrino oscillation model.
    }
\label{Fig:exclusion}
\end{figure}

\indent
These results can also be interpreted to the neutrino oscillation hypothesis.
The 90\% confidence level (C.L.) exclusion region derived from this analysis is shown in 
Fig.~\ref{Fig:exclusion}. 

Feldman–Cousins statistics~\cite{CITE:FC} are employed, 
incorporating fluctuations of the expected background and 
signal yields using Poisson statistics, with systematic uncertainties 
included at each
(sin$^2 2\theta$, $\Delta m^2$) point. 
The 90\% and 99\% C.L. allowed regions from the LSND experiment are also shown for comparison. 
The energy dependence of efficiencies on the IBD prompt energy and PSD are taken 
into account in constructing the exclusion curve.

\section{Summary}
\indent
We report the first results from a direct tests of LSND anomaly regarding 
the excess of $\bar{\nu}_{e}$ events in JSNS$^2$ using 2022 data. 
Two events are observed, in agreement with the background-only prediction of 
2.3$\pm$0.4.  
The current statistics do not allow a conclusive test of the LSND anomaly yet,
however this result is based 
on the data corresponding to 
approximately 7.2\% of the approved protons-on-target (POT) for JSNS$^2$. 
As of the end of 2025 the accumulated POT is roughly half of the planned amount.
Further improvements in the selection efficiency and an expansion of the fiducial 
volume by a factor of a few in total are also anticipated.
In addition, the baseline dependence 
of the anomaly will be investigated using two detectors in the experiment second phase, 
JSNS$^2$-II~\cite{CITE:JSNS2-II}, with increased POT. 
This two-detector configuration will reduce uncertainties in the expected signal rate by enabling a relative measurement as a function of baseline.

\begin{acknowledgments}
We deeply thank the J-PARC for their continuous support, 
especially for the MLF and the accelerator groups to provide 
an excellent environment for this experiment.
We acknowledge the support of the Ministry of Education, Culture, Sports, Science, and Technology (MEXT) and the JSPS grants-in-aid: No.\,16H06344, No.\,16H03967, No.\,23K13133, 
No.\,24K17074, No.\,20H05624 and No.\ 25H00649, Japan. This work is also supported by the National Research Foundation of Korea (NRF): No.\,2016R1A5A1004684, No.\,RS-2016-NR018633, No.\,17K1A3A7A09015973, No.\,017K1A3A7A09016426, No.\,2019R1A2C3004955, No.\,2016R1D1A3B02010606, No.\,017R1A2B4011200, No.\,2018R1D1A1B07050425, No.\,2020K1A3A7A09080133, No.\,020K1A3A7A09080114, No.\,2020R1I1A3066835, No.\,2021R1A2C1013661, 
No.\,NRF-2021R1C1C2003615, No.\,2021R1A6A1A03043957, No.\,2022R1A5A1030700, No.\,RS-2022-NR070836
No.\,RS-2023-00212787, No.\,2023R1A2C1006091, No.\,RS-2024-00416839, No.\,RS-2024-00442775 
and No.\,RS-2024-00416839. 
Our work has also been supported by a fund from the BK21 of the NRF. The University of Michigan gratefully acknowledges the support of the Heising-Simons Foundation. This work conducted at Brookhaven National Laboratory was supported by the U.S. Department of Energy under Contract DE-AC02-98CH10886. The work of the University of Sussex is supported by the Royal Society grant No.\,IESnR3n170385. We also thank the Daya Bay Collaboration for providing the Gd-LS, the RENO collaboration for providing the LS and PMTs, CIEMAT for providing the splitters, Drexel University for providing the FEE circuits and Tokyo Inst. Tech for providing FADC boards.
\end{acknowledgments}

\hspace{0.7cm}

\section*{DATA AVAILABILITY}
The data that support the findings of this article are not publicly available. The data are available from the authors upon reasonable request.

\nocite{*}

\bibliography{apssamp}

\end{document}